\documentclass[twocolumn,twoside,amsmath,amssymb,superscriptaddress,showpacs,nofootinbib,aps]{revtex4-2}
\usepackage{slashed}
\usepackage{amsmath}
\usepackage{amsthm}
\usepackage{subfigure}
\usepackage[utf8]{inputenc}
\usepackage{graphicx}
\usepackage{epsfig}
\usepackage{amssymb}
\usepackage{dcolumn}
\usepackage{bm}
\usepackage{color}
\usepackage{subfigure}
\usepackage{hyperref}
\usepackage{amssymb}
\newcommand{\ba}{\begin{array}}
\newcommand{\ea}{\end{array}}
\def\br{\begin{eqnarray}}
\def\er{\end{eqnarray}}
\def\be{\begin{equation}}
\def\ee{\end{equation}}

\def\({\left(}
\def\){\right)}

\def\<{\left\langle}
\def\>{\right\rangle}

\def\tt{\textnormal\tiny\textsc}
\definecolor{darkolivegreen}{rgb}{0.33, 0.42, 0.18}
\definecolor{babyblueeyes}{rgb}{0.63, 0.79, 0.95}
\begin{document}
\title{Composite scalar bosons masses: Effective potential versus  Bethe-Salpeter approach}
\author{A. Doff}
\email{agomes@utfpr.edu.br}
\affiliation{Universidade Tecnologica Federal do Parana - UTFPR - DAFIS, R. Doutor Washington Subtil Chueire, 330 - Jardim Carvalho,  84017-220
Ponta Grossa, PR, Brazil}

\date{\today}

\begin{abstract}
Ten years ago the $125$ GeV  Higgs resonance  was discovered at the LHC \cite{atlas,cms} ,  if this boson is a fundamental particle or a particle composed of new strongly interacting particles is still an open question. If this is a composite boson there are  still no signals of  other possible composite states of this scheme, a possible solution to this problem was recently discussed in Refs.\cite{el,lane}, where it is argued that the Higgs boson can be a  composite dilaton \cite{el}. In this work, considering an effective potential for composite operators  we verify that the potential responsible for a light composite scalar boson of $O(120)GeV$, behaves like  $\propto \Phi^4$  suggesting that if the Higgs boson is a composite scalar it  may  be a composite dilaton.
\end{abstract}

\maketitle
\section{Introduction}

\,\,\,\,\,\,\,\,\, The Higgs boson discovery was one of the major particle physics  breakthrough in the last decades. If this boson is composed by new strongly interacting particles is still an open question , composite scalar bosons appear in the context of Technicolor theories (TC) \cite{wei,sus,far}, which usually have a composition scale of order of $\Lambda_C \geq 1$TeV. Another possibility
 raised a few years ago is to have a Higgs arising as a composite pseudo-Goldstone boson (PGB) from the strongly interacting sector. In this case the Higgs mass is protected by an approximate global symmetry and is only generated via quantum effects,  models based on this approach are usually  called composite Higgs 
models\cite{bella}. 

Technicolor, was inspired in QCD,  to provide a natural and consistent quantum-field theoretic description of electroweak (EW) symmetry breaking,  without elementary scalar fields.  Crafting a realistic Technicolor model can be a very precise engineering problem.

\par Over the years different models have emerged considering  different approaches to obtain large mass anomalous dimension ($\gamma$), which in summary, can be classified as\\
 \noindent (i)  Quasi-conformal TC theories\cite{lane0,appel,aoki,appelquist,shro,kura}, where $\gamma(N_{TC}, n_F)$,  and it is possible to obtain an almost conformal behavior when the fermions are in the fundamental representation, $R=F$ , introducing a large number of TC fermions $n_F$. Nonetheless,  the cost of such procedure may be a large S parameter, in an the similar way an almost conformal TC theory can also be obtained when the fermions are in larger representations  that the fundamental one\cite{Foadi,Kou,Mads,Ari}. \\
\noindent (ii) Methods of lattice gauge theory,  as the authors in Refs.\cite{appelt1,appelt2,appelt3,appelt4} has demonstrated,   the conformal window for the $SU(3)$ gauge theory lies in the range $8 < n_f < 12$, and in this region is possible indeed have a slowly running coupling (or $\beta \approx 0$ ). \\
\noindent (iii)  Gauged Nambu-Jona-Lasinio models, in this case  the existence of an “effective” four-fermion interaction in TC dynamics\cite {yama1,yama2,yama22,yama3,mira3,yama4,takeuchi}, could also be responsible by large $\gamma$ values.

As proposed by Holdom\cite{holdom}, a  light composite scalar boson may be generated when the strong interaction theory (or TC) has a large mass anomalous dimension ($\gamma$), the  discussion of how these scenarios might lead to a light composite scalar boson  is presented  in Refs.\cite{rev1}. Assuming these different scenarios, calculations involving effective Higgs Lagrangians led to different predictions regarding the masses of composite scalar bosons.

The self-energy  of the new fermions (technifermions) responsible for the composite states, that are characterized by a large mass anomalous dimension $\gamma$, result in mass diagrams whose calculation do not scale with the naive dimensions. In this case, the self-energy $\Sigma_{\tt{F}}(p^2) $ at large momenta is proportional to

\be
\Sigma_{\tt{F}} (p^2) \propto \frac{\mu_{\tt{F}}^3}{p^2} (p^2/\mu_{\tt{F}}^2)^{\gamma/2} .
\label{eq0}
\ee
where $\mu_{\tt{F}}\sim O(1)TeV$ is the typical composition or dynamical fermion mass scale.

The absence of signals of a large scalar boson sector in the experimental data, as well as a possible explanation of a light Higgs boson, has been beautifully discussed recently in Refs.\cite{el,lane}. In that references it is argued that the Higgs boson, a Gildener-Weinberg dilaton \cite{gw}, could be a  composite dilaton \cite{el}.

Recently, considering constituents of same mass, we computed the composite scalar mass using Bethe-Salpeter equations (BSE) \cite{usBS}. The calculation was performed with the help of an ansatz in the form of Eq.(\ref{eq0}) for the constituent self-energy dependent on the mass anomalous dimension, the results  obtained indicate how the scalar mass $M_S$ can vary with the mass anomalous dimension($\gamma$) leading to $M_S(\gamma) = A(\gamma)\mu_{\tt{F}}$. At this point, we must emphasize that there are not in the literature other calculations  using the Bethe-Salpeter equations to address the question of a possible composite Higgs boson in Technicolor models. Usually, the solutions of the  Bethe-Salpeter  equations are used in hadronic physics, as, for example, in the study of meson spectroscopy\cite{Pm}. 

In this work, considering  effective potential for composite operators\cite{cjt, cs, us0} we extend the discussion started in Ref.\cite{usBS}, in particular we verify that the potential responsible for  a light composite scalar boson, $M_S \sim O(120) GeV$,  behaves like  $\propto \Phi^4$,  indicating the possibility that  the Higgs boson may in principle be a composite dilaton  as suggested in Refs.\cite{el,lane},  and its mass can be smaller than the composition scale as long as we have large anomalous dimensions. Moreover, we determine the mass obtained for the lightest pseudo scalar boson, $\Pi^{0} \sim \bar{N}N$
and  we retrieve the result described in Ref.\cite{us05},  assuming the Bethe-Salpeter equations.

This paper is organized as follows: In section II we determine the effective potential for composite operators  for the constituent self-energy considered in Ref.\cite{usBS}, in the section III we calculate the scalar composite scalar mass from the effective potential. With the help of Bethe-Salpeter equations (BSE), assuming the same conditions employed in the previous section, in the section IV we calculate the scalar boson mass and compare with the previous result. At the end of this section, we  determine $\Pi^{0}$ mass from the BSE equations. The section V contains our conclusions.

\begin{figure}[t]
\centering
\includegraphics[width=0.5\columnwidth]{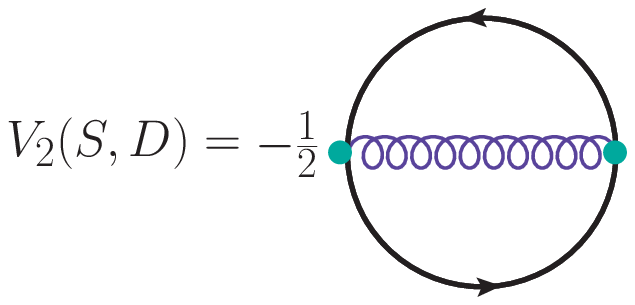}
\caption[dummy0]{Two-particle irreducible contribution to the vacuum
energy.}
\label{fig1}
\end{figure}


\section{The effective potential and the self-energy ansatz}

\subsection{The  effective potential for composite operators}

\par  The  effective potential for composite operators was proposed many years ago in Ref\cite{cjt}  by Cornwall, Jackiw and Tomboulis, as complementary references we suggest to the reader \cite{us0, us2},  where it is possible to find a detailed calculation of the  composite scalar masses considering   this method.  The effective action for composite operators $\overline{\Gamma}(G)$ is a function of the Green functions denoted by $G= G_n$, and is stationary with respect to variations of $ G_n$. The effective potential $V(G)$ is  defined by the following equation
\be 
V(G_n)\int d^4 x  = - \overline{\Gamma}(G)|_{ti},
\ee 
\noindent where $(ti:= translation\,\, invariant )$ , and $V(G)$ can be written in terms of the complete fermion (S) and gauge boson
(D) propagators, in the form
\br
&&\!\!\!\!\! V(S,D) = -iTr \left( \ln S_0^{-1} S - S_0^{-1} S + 1 \right)  + iV_2 (S,D),
\label{eq1}
\er
\noindent where in this expression  the complete fermion   propagator  is represented by  
\be 
S^{-1}(p) = {\slashed{p}} A(p^2) - B(p^2)\, .
\label{eq4}
\ee
\noindent whereas, the free propagator is $S_0= i/{\not\!p}$, and we shall assume $A(q^2)=1$ and $B(p^2)=\Sigma (p^2)$. 
In these equations, we are not considering contributions to the potential due to gauge and ghosts loops, 
we just are interested only in the fermionic bilinear condensation in the scalar channel, and will consider a non-abelian gauge theory, 
stronger than QCD, whose fermions form the composite scalar boson. In this case, we can consider the approximation where 
$D=D_0$ stands for the bare gauge boson propagator,  that in  Landau gauge, can be written as
\be 
g^{\mu\nu}D_{\mu\nu}(p-k) = \frac{3}{(p-k)^2} =  3G(p-k).
\label{D0}
\ee
\par In order to keep  the above equations in a compact form,  we are not writing the Lorentz and $SU(n)_{TC}$ indexes, as well as momentum integrals.  The last term in Eq.(\ref{eq1}), $V_2 (S,D)$, corresponds to the sum of all two-particle irreducible vacuum diagrams, shown in Fig.(\ref{fig1}), that can be analytically represented by
\be
i V_2 (S,D) = -\frac{i}{2}\,Tr (\Gamma S\Gamma S D),
\label{eq2}
\ee
\noindent where $\Gamma$ is the fermion proper vertex. An important property in the potential Eq.(\ref{eq1})  is that its minimization with respect to the complete propagators ($S$ or $D$) , $\delta V/\delta(S,D) = 0 $,  reproduce exactly the Schwinger-Dyson equations for the complete $S$ and $D$ propagators.

\par To obtain  an effective  Lagrangian for the composite scalar boson,  it is better to compute the vacuum energy density which is given by the effective potential calculated at minimum subtracted by its perturbative part, which does not contribute to dynamical mass generation
\be
\Omega_V = V(S,D) - V(S_0, D_0), 
\label{eq3}
\ee
After replacing  the Eqs.(\ref{eq1}), (\ref{eq4}) and (\ref{D0})  in Eq.(\ref{eq3}),  now  indicating the momenta integrals  we can  write 
\br
&&\Omega_V = i\int \frac{d^4p}{(2\pi)^4}Tr\left[\ln\left( 1 - \frac{\Sigma^2(p^2)}{p^2}\right) + \frac{\Sigma^2(p^2)}{p^2-\Sigma^2(p^2)}\right.\nonumber\\
&&\hspace*{0.6cm}\left.  + \frac{3C_2g^2(p^2)\Sigma^2(p^2)}{p^2(p^2 - \Sigma^2(p^2))} i\int^{p^2} \frac{d^4k}{(2\pi)^4}\frac{\Sigma^2(k^2)}{k^2(k^2 - \Sigma^2(k^2))}\right],\nonumber\\
\label{eq5}
\er
\noindent where  to obtain  the above equation we assumed the  angle approximation 
$$
\frac{g^2(p,k)}{(p-k)^2}=\frac{g^2(p^2)}{p^2}\Theta(p^2 - k^2) + \frac{g^2(k^2)}{k^2}\Theta(k^2 - p^2)
$$
and $\Theta$ is the Heaviside step function.  In the Euclidean space  we can write $\Omega_V$ as 
\be 
\Omega_V = \Omega_{V_A} + \Omega_{V_B}
\label{ve1}
\ee 
\noindent where 
\begin{widetext}
\br 
&&\hspace*{-1.3cm}\Omega_{V_A} = -\frac{N_{TC}n_F}{8\pi^2}\int p^2dp^2\left[\ln\left( 1 + \frac{\Sigma^2(p^2)}{p^2}\right)  - \frac{\Sigma^2(p^2)}{p^2+\Sigma^2(p^2)}\right], 
\label{ve2a}
\er 
and 
\br 
&&\Omega_{V_B} = +\frac{N_{TC}n_F}{8\pi^2}\int p^2dp^2\left[\frac{3C_2\alpha_{TC}(p^2)\Sigma^2(p^2)}{2\pi p^2(p^2 + \Sigma^2(p^2))}\int^{p^2} \frac{dk^2\Sigma^2(k^2)}{k^2(k^2 + \Sigma^2(k^2))}\right],
\label{ve2b}
\er 
\end{widetext}
\noindent in these equations, $N_{TC}$  is the number of technicolors (techniquarks are in the fundamental representation $R=F$ of $SU(N_{TC})$), and  $n_F=n_F(R)$ is the number of technifermions(F).

\subsection{The  self-energy  ansatz}

In usual calculations involving  the  Schwinger-Dyson equations, the self-energy $\Sigma (p^2)$ is obtained from the numerical solutions for the fermionic propagators. In this work we will assume the ansatz employed in \cite{usBS},  that is a function of the mass anomalous dimension $\gamma$, and has the form
\be
\Sigma (p^2) = \frac{\mu^3}{p^2+\mu^2} \( \frac{p^2+\mu^2}{\mu^2}\)^{\kappa} \, 
\label{eq6a}
\ee
where $\kappa = \gamma /2$. In this expression $\mu=\mu_{F}$ is the dynamically generated mass, and Eq.(\ref{eq6a}) behaves in the infrared region as  $\mu$ ,
as $\kappa \rightarrow 0$, Eq.(\ref{eq6a}) leads to self-energy roughly equal to $\mu^3/p^2$ in the asymptotic  region , which is the behavior predicted
by a standard operator product expansion (OPE) for the techniquark self-energy $\Sigma (p^2)$ for $\langle\bar{Q}Q\rangle \sim \mu^3$. 

\par However , when $\kappa \rightarrow 1$(or $\gamma \to 2)$ Eq.(\ref{eq6a}) can be written in the form
\be
\Sigma (p^2)\approx \mu \left[ 1+ \delta_1\ln\left[(p^2+\mu^2)/\mu^2 \right] \right]^{-\delta_2} \, ,
\label{eq7}
\ee
\noindent where to arrive at this form, $\delta_1$ and $\delta_2$ are obtained from $\gamma$ when expanded as a function of the running coupling $g^2(p^2)$.
In this case,  the self-energy stands for an extreme walking gauge theory, and the standard OPE prediction is modified by a large anomalous dimension $(\gamma)$, 
and the  self-energy behaves as $\mu \ln^{-\gamma}(p^2/\mu^2)$ asymptotically,  mapping the SDE possible solutions in the full Euclidean space, allow us to obtain an equation for the scalar mass $M_S$ as a function of $\gamma$.


\section{The effective  Lagrangian for composite scalar bosons}

\subsection{The kinetic term of the effective action}

\par In this section, we shall consider the problem of generating  one light composite scalar boson  in the context of the effective potential  assuming 
the self-energy ansatz  given by Eq.(\ref{eq6a}).  As pointed out recently in Ref.\cite{el,lane} a composite Gildener-Weinberg dilaton should result from an  effective potential that, at leading order(0), is proportional to
\be
V(\Phi)_{0} \propto \Phi^4 ,
\label{eq8}
\ee
where $\Phi$ is a composite effective field. A theory that generates a composite Higgs boson, where $\Phi \propto \bar{Q}Q$ and $\bar{Q}Q$ 
corresponds to a bound state of techniquarks Q, should naturally not contain a quadratic term in the effective potential.

\par The $\Sigma^2(p^2)/p^2$ term in the logarithm in Eq.(\ref{ve2a}) can be expanded, the contribution of the 
$\Sigma^2$ term in this equation eventually cancels out, leading to the absence of quadratic terms in the effective potential.
This cancellation is a consequence  of the fact that $\Sigma(p^2)$ obeys the linear homogeneous SDE for the fermion propagator\cite{us0, us2}.

\par  In order to obtain an expression for the effective Lagrangian of composite scalars bosons from Eq.(\ref{eq6a}), we will reconsider the approach described in Ref.\cite{cs} to determine a complete effective theory, including the kinetic term of the effective action. The fermionic propagator can be  described by a fermion bilinear  that  has the following operator expansion

\be
S(x,y) = \langle \Omega | T [\chi (x+\frac{1}{2} y) \psi (x-\frac{1}{2} y) ] |\Omega \rangle \,\,\, {}^{\,\, \sim}_{y \rightarrow 0}\,\,\,  C(y) \phi (x) , 
\label{eq9}
\ee
\noindent where in the above equation $C(y)$ is a $c$-number function, and $\phi (x)$ acts like a dynamical effective scalar field. Taking into account the structure of the real vacuum, where the propagator is a fermion bilinear and  $\Sigma$  depending on two momenta  $\Sigma (p,k)$, we can consider  the Fourier transform of above equation and write 
\be
\Sigma (p,k) \sim \phi (k) \Sigma (p) \, .
\label{eq10}
\ee
In the effective action, $\phi(x)$ can be seen as a variational parameter whose minimum will be indicated  by $\phi$, corresponding to the leading contribution of its expansion around $k=0$. A more detailed presentation of this  approach can be seen in Ref.\cite{cs}, as commented in Ref.\cite{us2}, depending on the theory dynamics, i.e. $\kappa \in [0, 1]$ in Eq.(\ref{eq6a}), when the self-energy decreases slowly with the momentum,  which corresponds to the case when $\kappa \to 1$  in 
Eq.(\ref{eq6a}),  the kinetic term  is important for the characterization of the effective Lagrangian. 

\begin{figure}[t]
\centering
\includegraphics[width=0.8\columnwidth]{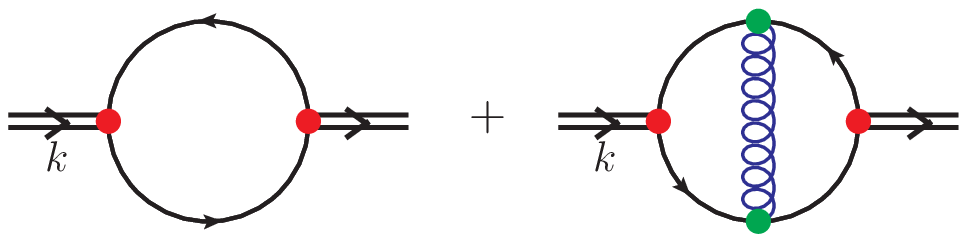}
\caption[dummy0]{ Diagrams contributing to the kinetic term in the effective Lagrangian.}
\label{fig2}
\end{figure}

The kinetic term is given by the polarization diagrams ($\Pi (k^2,\phi)$) shown in Fig.(\ref{fig2}), these diagrams are responsible in the effective Lagrangian for terms of the form
\be
\Omega_K =  \int d^4x \frac{1}{2}\partial_\mu \phi \partial^\mu \phi \, .
\label{eq11}
\ee
When the diagrams of Fig.(\ref{fig2}) are calculated it is possible to verify that this term will be multiplied by a non-trivial function ($Z(\kappa)$) , which has a dependency with  ($N_{TC}$), and the number of fermions ($n_F$),  and also $\gamma$, once we consider the anzatz for $\Sigma$ described by Eq.(\ref{eq6a}). This non-trivial function, that must normalize the scalar composite field $ \phi$, was obtained in the Ref.\cite{cs} and can be written as 
\be 
(Z(\kappa))^{-1} \approx \frac{N_{TC} n_{F}}{4\pi^2}\int dp^2\frac{(p^2)^2(\Sigma(p^2)/\mu)^2}{(p^2 + \mu^2)^3} \,\,\, ,
\label{eq14}
\ee 
 the index $\kappa = \frac{\gamma}{2}$ is the same one appearing in Eq.(\ref{eq6a}). 

\subsection{The effective Lagrangian }

\par Once we have characterized the kinetic term in the effective Lagrangian, now we can calculate the expansion for $\Sigma(p^2)/p^2 << 1$ in Eq.(\ref{ve1}), 
and write this equation  in terms of the variational field $\phi$  in the form \cite{us0,us2}
\br
&& \tilde{\Omega} \approx \int d^4x\left[\frac{1}{2Z^{(\kappa)}}\partial_\mu \phi \partial^\mu \phi -\lambda_{4(0)}\phi^4 + \lambda_{6(0)}\phi^6 - ...\right],\nonumber \\
\label{eq16}
\er
\noindent  where now 
\br 
&& \lambda_{4(0)} = \,\,\frac{N_{TC} n_{F}}{8\pi^2}\,\int dz \left(\,\,\,-\frac{3}{4}\frac{f^4(z)}{z} - f^2(z)\frac{df^2(z)}{dz}\,\,\, + \right. \nonumber\\
&& \left. + \frac{2}{3}\frac{f^6(z)}{z^2} + \frac{f^4(z)}{z^2}\frac{df^2(z)}{dz} -\frac{5}{8}\frac{f^8(z)}{z^3} - \frac{f^6(z)}{z^2}\frac{df^2(z)}{dz} + \right.\nonumber\\
&& \left. + \frac{3}{5}\frac{f^{10}(z)}{z^4} + \frac{f^8(z)}{z^3}\frac{df^2(z)}{dz}...\right)\nonumber\\
&& \lambda_{6(0)} = - \frac{\lambda_{4(0)}}{\mu^2},
\er

\noindent with $z=\frac{p^2}{\mu^2}$ and $f(z) =\frac{\Sigma(p^2)}{\mu}$. The effective coupling constants $\lambda_{4(0)}$ and $\lambda_{6(0)}$ are determined from Eqs.(\ref{ve1}) and (\ref{eq6a}). 
\par In Eq.(\ref{ve2b})  $C_2$ is the Casimir operator for fermions, and we consider $g^2(p^2) = g^2$ and  also the MAC hypothesis\cite{mac1,mac2} in this calculation, where $\frac{g^2C_2}{4\pi} \approx C_2\alpha \approx \frac{\pi}{3}$, that lead to 
\br 
&& \lambda_{4(0)}(\kappa)\approx \frac{N_{TC} n_{F}}{8\pi^2}\left[\frac{1}{2}  + \frac{3}{16\kappa - 16} + \frac{2}{3 (7-6\kappa)} \,+ \right. \nonumber\\
&& \hspace*{3cm}\left. -\frac{5}{8(10-8\kappa)} +\frac{3}{5 (13-10\kappa)} + \right. \nonumber\\
&& \hspace*{3cm}\left. -\frac{(\kappa-1) \left(68\kappa^2-172\kappa+109\right)}{(4\kappa-5) (6\kappa-7) (10\kappa-13)} \right] \nonumber \\
&& \lambda_{6(0)}(\kappa) \approx  -\frac{1}{\mu^2}\lambda_{4(0)}(\kappa). 
\label{eq17}
\er

The Eq.(\ref{eq16}) differs from a conventional scalar field Lagrangian by  the kinetic term $Z(\kappa)$, therefore, the final effective Lagrangian comes out when we normalize the scalar field  $\Phi$ according to  $\Phi\equiv [Z(\kappa)]^{-\frac{1}{2}}\phi$ , leading to the normalized effective Lagrangian $\tilde{\Omega}(\kappa)$
\br
\tilde{\Omega}(\kappa) = \int d^4x\left[\frac{1}{2} \partial_{\mu}\Phi\partial^{\mu}\Phi - \tilde{\lambda}_{4}(\kappa)\Phi^4 + \tilde{\lambda}_{6}(\kappa)\Phi^6 - ...\right], \nonumber\\
\label{eq19}
\er
\noindent  where,  assuming Eq.(\ref{eq6a}),  the normalization function $Z(\kappa)$ can be determined to be equal to
\be 
Z(\kappa) \approx \frac{4\pi^2}{N_{TC} n_{F}}\left(2 -2\kappa\right).
\label{eq15}
\ee 
\par The coupling constants $\lambda_{4(0)}$, $\lambda_{6(0)}$ indicated in Eq.(\ref{eq17}), were replaced by the respective normalized ones $\tilde{\lambda}_{4}(\kappa) \equiv Z^{2}(\kappa )\lambda_{4(0)}$ and $\tilde{\lambda}_{6}(\kappa) \equiv Z^{3}(\kappa )\lambda_{6(0)}$. Therefore, the scalar mass $M^2_{S}(\kappa)$  can be determined from the effective potential Eq.(\ref{eq19}) at the minimum from  
\be
M^2_{S}(\kappa) = \frac{\partial^2\tilde{\Omega}(\kappa)}{\partial \Phi^2}{|}_{{}_{\Phi=\Phi_{min}}}\!\!\!\!\!\approx 2\frac{ (\tilde{\lambda}_{4}(\kappa))^2}{\tilde{\lambda}_{6}(\kappa)} \approx 2Z(\kappa)\lambda_{4(0)}(\kappa)\mu^2 .
\label{eq20}
\ee
\par   Note that the effective potential depends on the TC fermionic representation $R$, the product $(N_{TC}\times n_F(R))$, and the anomalous dimension $\gamma=\gamma(N_{TC},n_F(R))$. In order to present numerical results as a function of $\gamma$ (which is also a function of $N_{TC}$ and $n_F(R)$), we will simply assume that
 TC is not so different from QCD and adopt $N_{TC}=3$ and $n_F=5$[33], expecting that we can still present the variation of the TC masses with $\gamma$  when determined from the effective potential. At this point we must point out that
$$
M^2_{S}(\kappa)/\mu^2 \approx 2Z(\kappa)\lambda_{4(0)}(\kappa) \propto \frac{\not{\!\!N}_{TC} \not{\!n}_{F}}{\not{\!\!N}_{TC} \not{\!n}_{F}}f(\kappa)  \propto f(\kappa)
$$
\noindent i.e. the dependence of $M^2_{S}(\gamma,N_{TC}, n_F(R)) = M^2_{S}(\gamma) $ results only from the anzatz dependency, Eq.(\ref{eq6a}), with  $\gamma$. 

\par In usual perturbative QCD, we can expect $\gamma \sim 0$, however, as we commented large mass anomalous dimensions $\gamma$  could be obtained in scenarios as listed (i)-(iii). 

\par An increase of $n_F$, for example as the authors in Refs.\cite{appelt1,appelt2,appelt3,appelt4} have demonstrated the conformal window for the $SU(3)$ gauge theory lies in the range $8 < n_f < 12$, or the consideration of higher TF representations $R > F$\cite{Foadi,Kou,Mads,Ari}, would lead to large values to $\gamma$ needed to produce changes in the TC mass function. 
In calculations involving the  Eq.(\ref{eq6a}),  we will consider $\gamma$  as an adjustable parameter, and if the Higgs boson is a composite particle, a realistic TC model could probably  be characterized by one of these  possible  scenarios.


\section{The TC scalar mass: BSE versus effective potential approach}

\subsection{The $SU(N)_{TC}$ scalar masses}

\par In this section, we will reconsider the approach employed in Ref.\cite{usBS} for the characterization of the TC scalar and pseudo scalar
masses using the Bethe-Salpeter equations (BSE). 

\par   We will assume  that  (BSE)  equations described in Refs.\cite{jm1,jm2,jm3}, with the approximations discussed in\cite{usBS},  with the difference that in this work we will consider  $G_{\rho\nu} (k-q)$ given by Eq.(\ref{D0}), in order to compare the results obtained with the effective potential. 

The Bethe-Salpeter equation can be written as
\br
&&\chi^{tc} (p,q) = -\imath \int \frac{d^4k}{(2\pi)^4}S(q+\alpha p)K_{\rho\nu}(p,k,q) S(q-\beta p) ,\nonumber \\
&& K_{\rho\nu}(p,k,q) = \gamma_\rho \chi^{tc} (p,k)\gamma_\nu G_{\rho\nu}(k-q),
\label{eq21}
\er
\noindent where in the above expression $\chi^{tc}$ is the technicolor(tc) BS wave function,  $\alpha$ and $\beta$ characterize the fraction of momentum carried by the constituents,  with $\alpha + \beta = 1$. 

\par In the calculation, we will consider that each constituent carries half of the momentum , i.e. $\alpha = \beta = 1/2$. In the Eq.(\ref{eq21}) the fermion propagator is given by (\ref{eq4}),  and the BSE solution appear as an eigenvalue problem for $p^2=M^2$, where $M$ is the bound state mass.

\par In the equation(\ref{eq21}) the variables are $p,q,k$,  $k$ is integrated, and we remain with an equation in $q$ that will have a solution for $p^2=M^2$. To solve this integral equation, to scalar channel we can project  $\chi^{tc} (p,q) = \chi^{tc}_S (p,q)$  into four coupled homogeneous integral equations
given by 

\be
\chi^{tc}_S (p,q)=\chi_{S0}+\slashed{p}\chi_{S1}+\slashed{q}\chi_{S2}+
[\slashed{p},\slashed{q}]\chi_{S3},
\label{eq6}
\ee 
which are functions of $p^2$, $q^2$ and $p.q=pqcos\theta$. It is possible to expand $\chi^{tc}(p,q)$ in terms of  Tschebyshev polynomials, and these equations can be truncated at a given order determined by the relative size of the next-order functions.  A satisfactory solution can be obtained by keeping only some terms, like $\chi_{S(0,1)}^{(0)},  \chi_{S(1)}^{(0,1)},  \chi_{S(2)}^{(0,1)}$. 

\par In addition to the above considerations we will consider  constituents of same mass $m = m_a = m_b$,  with  $m = \Sigma (x+\frac{1}{4}p^2)$, where  $x=q^2/\mu^2$and $\mu =\mu_{tc}$  is   the characteristic mass scale of the binding forces(TC).

\begin{figure}[t]
\centering
\includegraphics[width=0.8\columnwidth]{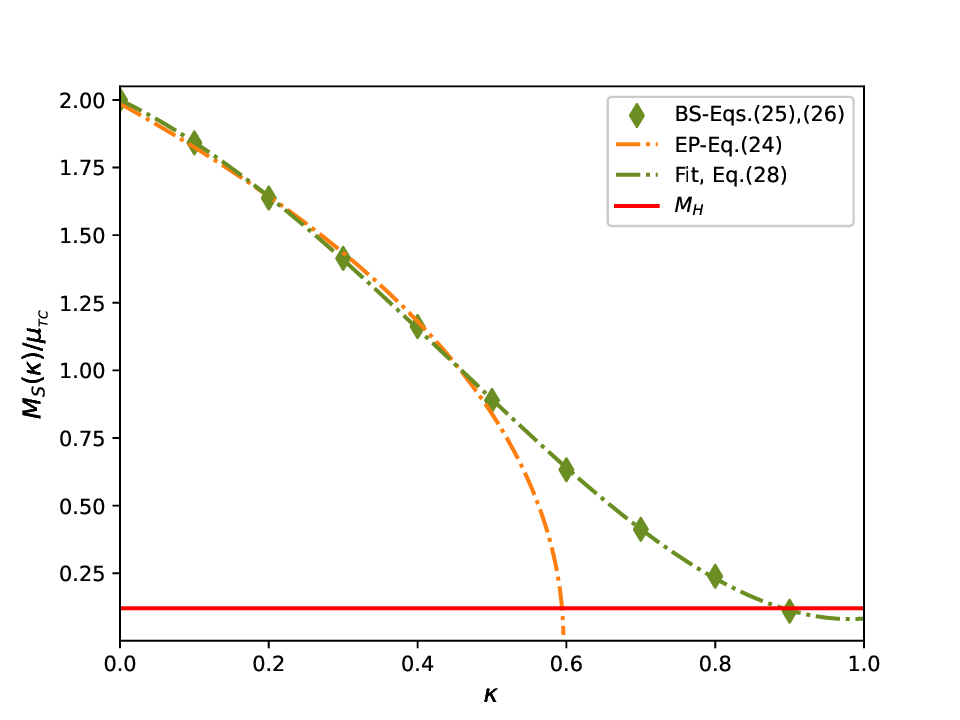}
\caption[dummy0]{Scalar masses $M_S(\kappa)$ considering different approaches. The contextualization of the curves behavior are described in the text. }
\label{fig4}
\end{figure}


\par The procedure for determining $M_{S}(\kappa)$ is the same one described in Ref.\cite{usBS}. In Fig.3 we present the behavior 
obtained for $M_{S}(\kappa)/\mu_{tc}$ considering the anzats for $\Sigma(p^2)$ given by Eq.(\ref{eq6a}), assuming Eq.(\ref{eq20}) and the corresponding result obtained  from Eq.(\ref{eq21}). As in Eq.(\ref{eq20}) the scalar mass depicted in the Fig.3 is just a function of $\gamma$, which is an adjustable parameter.

\par In this figure we normalize ours results for $M_S$ in terms of
\be 
M_S = 2\mu_{tc}. 
\label{enorm}
\ee
associated to a negligible $\gamma$. 

\par The choice of this normalization is based on the result described in Ref.\cite{DS}, where Delbourgo and Scadron  verified analytically with the help of the homogeneous BSE equations, that the sigma meson  mass is given by  $m_{\sigma} = 2\mu_{dyn}$.  We can use this result obtained for QCD to determine, by appropriate rescaling, the behavior of $M_S$ in TC models, where  $\mu_{tc}= 1TeV$. 

\par In the Fig.3, the dot-dashed line in orange matches the results obtained from Eq.(\ref{eq20}),  the points denoted in ({\color{darkolivegreen} $\blacklozenge$}) represent the BSE numerical solutions obtained for $M_{S}(\kappa)/\mu_{tc}$. The  line, in red,  corresponds to the radius $M_H/\mu_{tc}$ for  comparison with the $M_{S}(\kappa)/\mu_{tc}$  results.

\par In the dot-dashed line in olive we show the fit for $M_{S}(\kappa)/\mu_{tc}$ data with $R^2 =0.99997$ which corresponds to 
\br
&& S(\kappa) = \frac{M_S(\kappa)}{\mu_{tc}} = 2 - 1.3161\kappa  - 2.65806\kappa^2  \nonumber\\
&& \hspace*{2.4cm}                          + 1.38845\kappa^3 + 0.66806\kappa^4.
\label{eqfit1}
\er 

\par The expansion considered in  Eq.(19) is not sensitive to the region of low momenta, which is captured by the BS equations, so that at $\gamma \approx  1$ the curves start to show a different behavior. 

\par The comparison of the behavior exhibited for $M_{S}(\kappa)/\mu_{F}$, obtained with the different approaches, suggests that the potential responsible for generating the composite light scalar behaves like  $\propto \Phi^4$,  being characterized by $\gamma \sim O(1.2-1.8)$ indicating that the composite scalar boson
$\Phi$ seems to behave  like a dilaton as suggested in\cite{el,lane}.

\subsection{TC pseudo scalar masses}

\par Let us suppose that the TC group is not so different from QCD where there are many pseudo Goldstone bosons (or technipions) resulting from the chiral symmetry breaking of the technicolor theory.

\par These technipions, besides the ones absorbed by the W's and Z gauge bosons, can be classified for example according to\cite{us04}:
\par\noindent (a) Charged and neutral color singlets:
\br 
&&\Pi^{\pm} \sim \bar{U}^iD_i + \bar{D}^iU_i - 3\bar{N}E \nonumber\\
&&\Pi^{0}  \sim \bar{U}^iU_i + \bar{D}^iD_i - 3(\bar{N}N + \bar{E}E),  
\er
\noindent where $(i)$ denote de number of  TC flavours.
\par\noindent (b) Colored triplets: 
\be
\Pi^{3} \sim \bar{N}U^{a}_i + \bar{E}U^{a}_i,
\ee 
\par\noindent (c) Colored octets: 
\be
\Pi^{8} \sim \bar{U}^{a'}_iU^{a}_i + \bar{D}^{a'}_iD^{a}i,
\ee
\noindent in the above expressions $(a, a')$  denote a color index.
\par The heaviest pseudo Goldstone boson carries color once they have large radiative corrections from QCD, while others may have only
electroweak corrections to their masses. 

\par The lightest technifermion will be the neutral one (N), and the lightest pseudo Goldstone $\Pi^{0} \sim \bar{N}N$, and we assumed that such neutral boson is composed by (N) technifermions. From this point we can determine $\Pi^{0}$ mass, $M_{\Pi^{0}}$,  from the BSE equations. 

\par  For pseudo scalar components  the projection of $\chi^{tc} (p,q) = \chi^{tc}_{P}(p,q)$ is  given by
\be
\chi^{tc}_P (p,q)= \gamma_5(\chi_{P0}+\slashed{p}\chi_{P1}+\slashed{q}\chi_{P2}+
[\slashed{p},\slashed{q}]\chi_{P3} \,),
\label{eqps}
\ee 
\noindent the  components of Eq.(\ref{eqps}), $\chi_{P(i)}^{(0,1)}$ for $i=0..3$,   are listed in Appendix A. Assuming  the Eqs.(\ref{eq21}) and (\ref{eqps}),  in Fig.4 we present the behavior for $M_{\Pi^{0}}(\kappa)/\mu_{tc}$ compared to $M_{S}(\kappa)/\mu_{tc}$, where  we again normalize our results  in terms of Eq.(\ref{enorm}). 

\par In this figure the dot-dashed line in olive correspond to the fit of $M_{S}(\kappa)/\mu_{tc}$ given by Eq.(\ref{eqfit1}), the points indicated by ({\color{babyblueeyes} $\blacktriangle$}) in Fig.(3) represent the BSE numerical solutions obtained for $P(\kappa) = M_{\Pi^{0}}(\kappa)/\mu_{tc}$.

\par The dot-dashed light blue line represents the fit for $M_{\Pi^{0}}(\kappa)/\mu_{tc}$ data with $R^2 =0.999997$,  which corresponds to 
\br
&& P(\kappa) = M_{\Pi^{0}}(\kappa)/\mu_{tc} = 2 - 1.88021\kappa + 0.0626412\kappa^2 \nonumber \\
&&\hspace*{3.15cm}                           - 0.792769\kappa^3 + 0.780151\kappa^4. 
\label{eqfit2}
\er 

\par Therefore, from Eqs.(\ref{eqfit1}) and (\ref{eqfit2}), we determine the radius for $M_S(\kappa) = M_H$ as 
\be 
\frac{M_{PS}}{M_S} = 2.47,
\ee 
\noindent which represents the following lower bound on the lightest pseudo scalar mass $M_{\Pi^{0}} \approx 309 GeV$. This result confirms the estimate presented in Ref.\cite{us05}, which corresponds to
\be 
M_{\Pi^{0}} \approx (200 - 460)GeV. 
\ee
\noindent where we had also assumed that such neutral boson is solely composed by N technifermions. 

\begin{figure}[t]
\centering
\includegraphics[width=0.8\columnwidth]{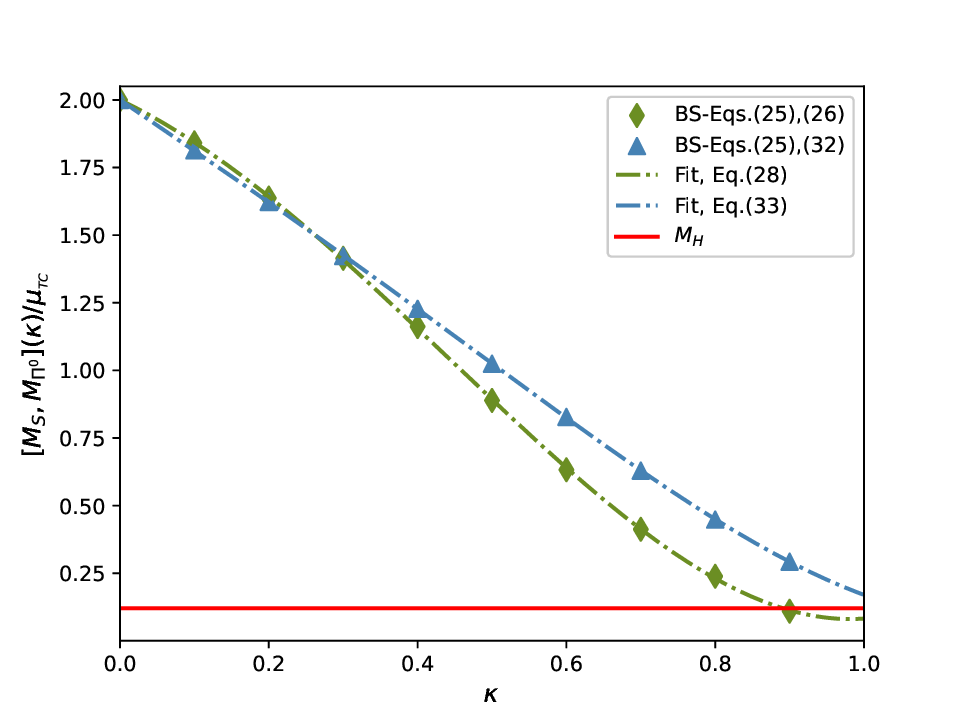}
\caption[dummy0]{Pseudoscalar masses $M_{\Pi^0}(\kappa)/\mu_{tc}$(curve in  light blue) and scalar masses depicted in Fig.(3). The contextualization of the curves behavior are described in the text. }
\label{fig4}
\end{figure}
\section{Conclusions}

In this work, we verify that the effective potential responsible for $M_S \sim O(120)  GeV$  behaves like  $\propto \Phi^4$ at leading order. We also considered the comparison between two different approaches to obtain $M_S(\gamma)$, i.e. the effective potential for composite operators and Bethe-Salpeter equations. These results corroborate with the hypothesis that if the Higgs boson is a composite scalar,  it may be a composite dilaton as suggested  in Refs.\cite{el,lane}.

This result is displayed in Fig.(3),  the curves obtained with different approaches, considering $G(k-q)$ given by Eq.(\ref{D0}), overlap exactly up to $\gamma \approx 1$. As we commented, the expansion considered in the  Eq.(19) is not sensitive to the region of low momenta, which is captured by the BS equations, so that at $\gamma \approx  1$ the curves start to show different behaviors.

In the last section of this work, we include the determination of the lightest pseudo scalar mass $M_{\Pi^{0}} \approx 309 GeV$, that confirms the estimate presented in Ref.\cite{us05}. Charged and colored technifermions will not only have larger masses than the neutral technifermion(N), but also more radiative corrections to their masses, and we can expect even larger masses for these colored and charged pseudo scalar bosons.   

According to the discussion presented in Ref.\cite{us2}, we still have other contributions to the effective Lagrangian, given by Eq.(\ref{eq16}). These contributions are the ones coming from ordinary massive quarks and leptons that couple to the composite scalar boson $\Phi$. These contributions will be dominated by the heaviest fermion, the quark top,  and will generate terms of order $\Phi^3$ and $\Phi^4$. 

However, as we verified in this work the  $\Phi^4$ contributions to $\tilde{\Omega}$ from massive fermions can be disregarded, the only exception is the contribution to $\Phi^3$ term,  which is small but introduce some effect in the scalar mass calculation. If the Higgs boson is a composite particle  it is still possible that its constituents are bounded by a non-Abelian gauge strong interaction, and we believe that combining different approaches can be useful for characterizing the properties of this possible composite state.

\appendix
\section{The pseudo scalar components  $\chi_{P(i)}^{(0,1)}$ in the BS equations}

\par Assuming that  each constituent, the (N) techniquarks, carries half of the momentum $\alpha = \beta = 1/2$ , the different components of Eq.(\ref{eqps}), $\chi_{P(i)}^{(0,1)}$ for $i=0..3$, are listed  in the sequence
\be
\chi^{(0)}_{P0}(x,p^2)=3[(x-\frac{1}{4}p^2+m^2) J_1] I_{P0} + 
\Delta\chi^{(0)}_{P0} \, ,
\label{eq9a}
\ee
where 
\be
I_{P0} = \frac{2}{3\pi}\int dy y \chi^{(0)}_{P0} K_1 \, ,
\ee
with $y=k^2/\Lambda^2$ and 
\be
K_1(x,y)=\frac{3}{16\pi^2} \int d\theta sen^2 \theta G(x,y,cos \theta) \, ,
\ee
\be
J_1= \frac{2}{\pi} \int_0^\pi d\theta\frac{sen^2 \theta}{D(p^2,q^2, pq cos\theta )}
\label{eq12a}
\ee
 and in the Eq.(\ref{eq12a}), we have 
\be
D(p^2,q^2, pq cos\theta )=\{ (q+\frac{1}{2}p)^2+m^2 \}\{ (q-\frac{1}{2}p)^2+m^2 \}.
\ee
\par Considering Taylor's series expansion  of  $(q+\frac{1}{2}p)^2+m^2$ and $(q-\frac{1}{2}p)^2+m^2$ , 
keeping the first-order  derivative terms for $m$,  we have that the function $J_1$ can be written  as 
\br
J_1 = \frac{2}{c_1c_4 + c_2c_3}\left[\frac{c_2}{D_1}+ \frac{c_4}{D_2} + d_1\left(\frac{c_1}{D_1} -\frac{c_3}{D_2}\right)\right]
\er
where, in our approximation  with $\alpha = \beta = 1/2$, we obtain
$$ c_1=c_3 = x+\frac{1}{4} p^2 + m^2$$
$$ c_2=c_4 = 1+2m m^\prime$$
and $m^\prime$ is the derivative of $m$ with respect to the momentum. In addition, as a consequence of $\alpha=\beta$ 
$$ d_1=0$$
$$ D_1=D_2= c_1+\sqrt{c_1^2-p^2xc_2^2}. $$
\noindent In Eq.(\ref{eq9a}) , the term $\Delta\chi^{(0)}_{P0}$ stands for corrections to the leading-order results for $\chi^{(0)}_{P0}$,
that correspond to  $\chi^{(0,1)}_{P1}$ , $\chi^{(0,1)}_{P2}$ and $\chi^{(0,1)}_{P3}$. 
\par With the approximations considered, we have
\br
&&\Delta\chi^{(0)}_{P0} = -\frac{2}{3\pi} m p^2 J_1 \int dy y \chi_{P1}^{(0)}(K_1 + 2yK_3) +  \nonumber\\
&&\hspace*{1.4cm}-\frac{4}{9\pi} p^2 J_3\int dy y \chi_{P1}^{(0)}(3K_1 - 4yK_3) + \nonumber\\
&&\hspace*{1.4cm}+\frac{2}{3\pi}  p^2(J_1 - J_3)h(K_6,K_3,x,y) \nonumber\\
&&\hspace*{1.4cm}+\frac{2}{3\pi}\left[(x - \frac{1}{4}p^2+ m^2)((J_1 - 4J_3)p^2\right]\nonumber\\
&&\hspace*{1.4cm} \times h(K_7,K_1,y),
\label{eq13a}
\er
\noindent where we define,
\br
&& h(K_6,K_3,x,y) = \int dy y\chi_{P3}^{(0)}\left(2\sqrt{xy}K_6 -\frac{8}{3}xy K_3\right) \nonumber \\
&& h(K_7,K_1,y) = \int dy y^2 \chi_{P0}^{(2)}\left(\frac{4}{3} K_7 - K_1 \right). 
\er 
\par In the equation above, the lowest order terms $\chi_{P(1-3)}^{(0)} $ are given by 
\begin{widetext}
\be 
\chi_{P1}^{(0)} = \frac{mJ_1}{J_1(x - \frac{1}{4}p^2+ m^2)}\chi^{(0)}_{P0}\,\,,\,\,\chi_{P2}^{(0)} = 0\,\,,\,\, \chi_{P3}^{(0)} = \frac{1}{2}\frac{J_1}{J_1(x - \frac{1}{4}p^2+ m^2)}\chi^{(0)}_{P0}. 
\ee 
\end{widetext}
\par While the higher order term, $\chi_{P0}^{(2)}$ is given by
\be 
\chi_{P0}^{(2)} = \frac{1}{xp^2}\frac{4(x - \frac{1}{4}p^2+ m^2)J_3 - J_1}{J_1(x - \frac{1}{4}p^2+ m^2)}\chi^{(0)}_{P0}. 
\label{eq14a}
\ee 
\par We are dealing with scalars(S) and pseudo scalars(PS)  bosons  with equal mass constituents  and in this case 
we have simpler equations,  compared to Ref.\cite{jm2}, that correspond to
\be
J_2 = 0\,\,\,,\,\,\,J_3= \frac{1}{D_1^2}.
\ee
\par In the Eqs.(\ref{eq13a}-\ref{eq14a}), the kernels $K_i(x,y)$, with $i=3,6,7$ are given by 
\be
K_3(x,y)=\frac{3}{16\pi^2} \int d\theta \frac{sen^4 \theta}{x+y-2\sqrt{xy} cos\theta} G(x,y,cos\theta) ,
\label{eq20a}
\ee
\be
K_6(x,y) = \frac{3}{16\pi^2} \int d\theta sen^2 \theta cos\theta G(x,y,cos\theta) ,
\label{eq21a}
\ee
\be
K_7(x,y) = \frac{3}{16\pi^2} \int d\theta sen^4\theta G(x,y,cos\theta) .
\label{eq22a}
\ee
\par As in the determination of $M_{S}(\kappa)$, to obtain $M_{PS}(\kappa)$ we consider  $G_{\rho\nu}$  in the Landau gauge given by Eq.(\ref{D0}), in addition we also consider the MAC hypothesis\cite{mac1,mac2}. Therefore, in this equation  $G(x,y,cos\theta) = \mu^2 G(k-q)^2$, with $G(k-q)$ given by Eq.(\ref{D0}), that leads to 
\be 
\mu^2 G(k-q) = \frac{\mu^2}{p^2 + k^2 - 2pqcos\theta} = G(x,y,cos\theta). 
\ee 
\vspace*{-0.7cm}
\section*{Acknowledgments}
I would like to thank A. A. Natale for reading the manuscript and for useful discussions. This research  was  partially supported by the Conselho Nacional de Desenvolvimento Cient\'{\i}fico e Tecnol\'ogico (CNPq) under the grant 310015/2020-0 (A.D.).

\end{document}